\documentclass{moriond}

\def\be{\begin{equation}}
\def\ee{\end{equation}}
\def\bea{\begin{eqnarray}}
\def\eea{\end{eqnarray}}

\def\PH     {\ensuremath{\mathrm{H}}}
\def\pl     {\ensuremath{\mathrm{l}}}

\def\PV{\ensuremath{\mathrm{V}}} 
\def\PZ{\ensuremath{\mathrm{Z}}} 
\def\tautau{\ensuremath{\tau^{+}\tau^{-}}}
\def\WW{\ensuremath{\mathrm{W}^{+}\mathrm{W}^{-}}} 
\def\WWWW{\ensuremath{\WW\WW}} 
\def\ggWW{\ensuremath{\WW\gamma\gamma}} 
\def\bbbar{\ensuremath{\mathrm{b}\bar{\mathrm{b}}}}
\def\bbbb{\ensuremath{\bbbar\bbbar}}
\def\bbgg{\ensuremath{\bbbar\gamma\gamma}}
\def\ggbb{\ensuremath{\gamma\gamma\bbbar}}
\def\bbtt{\ensuremath{\bbbar\tautau}}
\def\bbVV{\ensuremath{\bbbar\PV\PV}}

\def\lbdHHH{\ensuremath{\lambda_{\PH\PH\PH}}}

\def\mh       {\ensuremath{m_\PH}}

\def\kapl{\ensuremath{\kappa_{\lambda}}}
\def\k2v{\ensuremath{\kappa_{2\PV}}}
\def\ifb{\ensuremath{\mathrm{fb}^{-1}}}
\def\ttH{\ensuremath{\mathrm{t} \bar{\mathrm{t}} \mathrm{H}}}

\newcommand{\Photo}{\includegraphics[height=35mm]{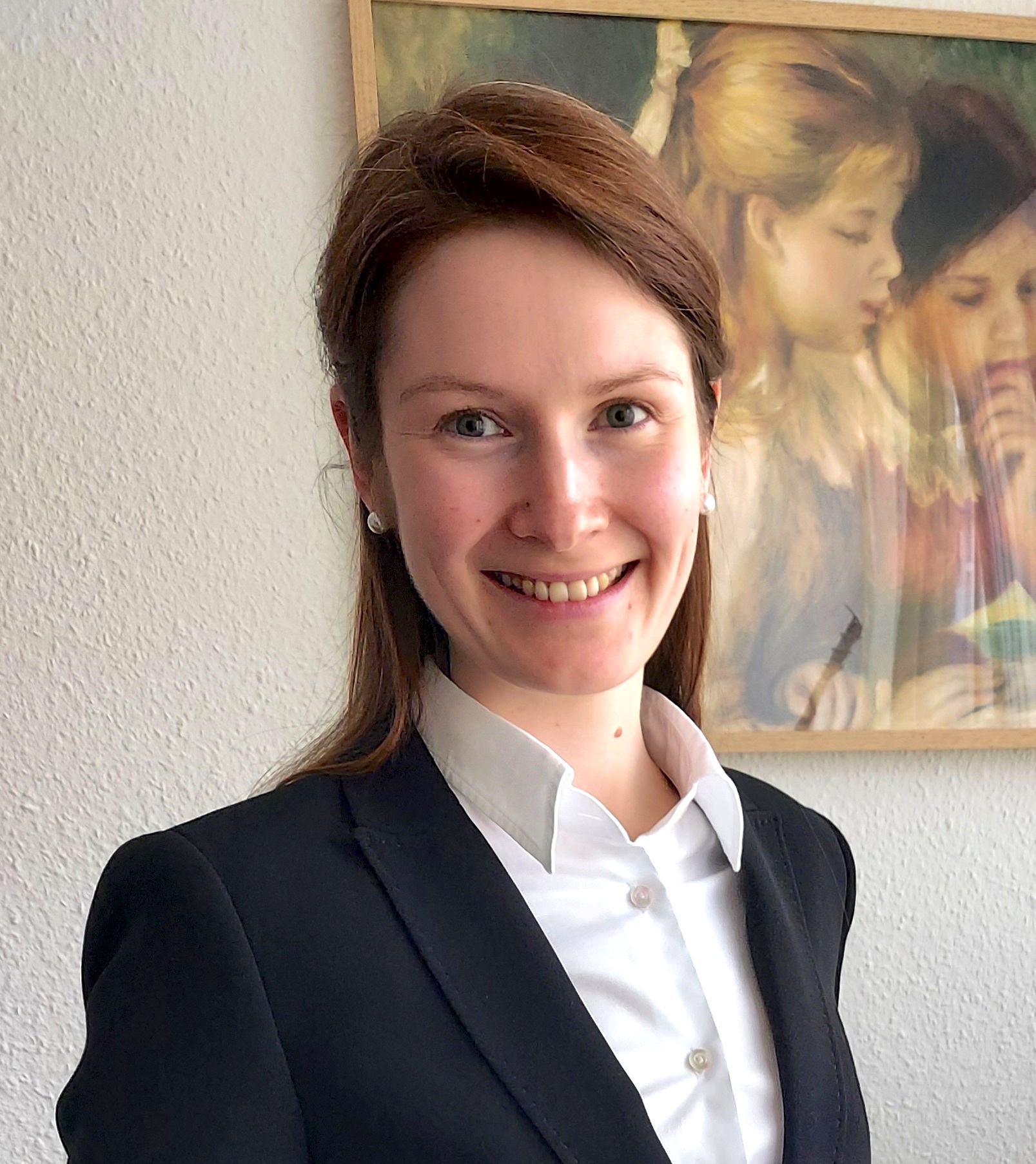}}

\begin{document}
\vspace*{4cm}
\title{Di-Higgs searches at the LHC}

\author{ Nadezda Chernyavskaya \\On behalf of the ATLAS and CMS Collaborations }

\address{CERN, Experimental Physics Department, \\Espl. des Particules 1, 1211, Switzerland}

\maketitle\abstracts{
An overview of the recent searches for Higgs boson pair production at the LHC was presented. The searches were based on approximately 140 \ifb of data collected by the ATLAS and CMS experiments in proton-proton collisions at $\sqrt{s}$ = 13 TeV. With respect to the previous searches, analysis techniques were significantly improved, and new signatures and decays channels were explored. }

\footnotetext{\textsf{\copyright~2021 CERN for the benefit of the ATLAS and CMS Collaborations.  CC-BY-4.0 license.}}

\section{Introduction}

Since the discovery of the Higgs boson (H) in 2012~\cite{Chatrchyan:2012ufa,Chatrchyan:2013lba,Aad:2012tfa}, enormous progress has been made in measuring its properties. The mass of the Higgs boson was measured to a per-mille precision level, and its interactions with vector bosons and fermions were established in multiple decay channels. However, the Higgs boson self-interaction and the energy potential of the Higgs field are not yet measured experimentally. 

A broad physics program is covered in the searches for di-Higgs (HH) production at the LHC. Measuring HH production is the only direct way to access the Higgs boson trilinear self-coupling \lbdHHH. In the standard model (SM) the Higgs boson self-coupling and the structure of the scalar Higgs field potential are fully predicted in terms of the Higgs boson mass and the Fermi coupling constant. Any deviation from the predicted shape of the scalar potential can have fundamental implications on our understanding of the origin and the fate of the universe. Therefore, measuring the Higgs boson’s trilinear self-coupling is of particular importance as it allows to characterize the Higgs field potential. 

At the LHC, HH production is a very rare process with a total production cross section roughly three orders of magnitude smaller than the one of the single H production. However, the direct relation to the scalar potential makes HH production very sensitive to contributions from physics beyond the SM (BSM). Some BSM models predict new spin-0 and spin-2 resonances with masses varying from 250 GeV to a few TeV, and which have a sizeable branching fraction (BR) to a pair of Higgs bosons. In addition, the effects of BSM physics in the quantum loops or through modification of the SM Higgs boson couplings could significantly enhance the nonresonant HH cross section and change the kinematic properties of the HH signal. 

This document reviews searches for HH production by the ATLAS~\cite{Aad:2008zzm} and CMS~\cite{Chatrchyan:2008aa} Collaborations. The most recent results using the full Run-II dataset of $\sqrt{s}$ = 13 TeV pp collisions with an integrated luminosity of about 140 \ifb are presented, and future prospects are discussed.

\section{Overview of HH searches}\label{sec:overvew}
Measuring the HH production requires reconstructing the decay products of the Higgs bosons. There is a rich variety of final states to explore at the LHC, however to keep the total branching fraction high the pursued searches largely rely on the final states where at least one of the Higgs bosons decays to a pair of b quarks. Different final states are complementary, and present a trade-off between the total branching fraction and background contamination associated to a particular final state. HH production dominantly occurs via gluon-gluon
fusion (ggF), with the cross section predicted in the SM of $31.05^{+1.41}_{-1.99}$ fb~\cite{Grazzini:2018bsd,Borowka:2016ehy}, calculated at
next-to-next-to-leading order (NLO) with the resummation at next-to-next-to-leading-logarithm
accuracy and including top-quark mass effects at NLO. The second largest production mode, vector boson fusion (VBF), has a cross section of only $1.726\pm 0.036$ fb~\cite{Dreyer:2018qbw,Liu-Sheng:2014gxa}, calculated at next-to-NNLO, but gives a unique access to the coupling between a vector boson
pair and a Higgs boson pair (VVHH). In addition, small deviations of the VVHH coupling with respect to the SM value lead to very large enhancements in the cross section allowing to constrain this coupling already with Run-II data of the LHC.

The ATLAS and CMS Collaborations performed searches for HH production via ggF in the \bbbb, \bbgg, \bbtt, \bbVV, \WWWW, $\ggWW$ final states at $\sqrt{s} = 13$ TeV using part of the Run-II data sample with an integrated luminosity of about 40 \ifb collected in 2016. Statistical combinations of these
searches~\cite{Aad:2019uzh,Sirunyan:2018ayu} were performed, and observed (expected) upper limits at 95\% confidence level
(CL) were set on the signal strength of HH production with respect to the SM expectation which correspond to 6.9 (10.0) in ATLAS and 22.5 (12.8) in CMS. The Higgs boson self-coupling was also constrained, and the 95\% CL observed (expected) allowed interval for the coupling modifier $\kapl$, defined as a ratio of the measured $\lbdHHH$ value to the value predicted in the SM, is $-5.0 < \kapl < 12.0$ ($-5.8 < \kapl < 12.0$) in ATLAS, and $-11.8 < \kapl < 18.8$ ($-7.1 < \kapl < 13.6$) in CMS. The new searches using full Run-II data sample are presented in the next Section~\ref{sec:run2}.

\section{Searches for HH production with full LHC Run-II data }\label{sec:run2}

\subsection{Nonresonant HH production in the final state with 4 leptons and 2 b jets}
The first result of the search for nonresonant HH production where
one Higgs boson decays to a Z boson pair which decays to 4 leptons (\pl), where $\pl$ is either an electron or a muon, and the other to a pair of b
jets is presented by the CMS Collaboration~\cite{CMS:2020gxr}. The final state has a very small branching fraction of only 0.014\% but has small backgrounds and exhibits a clear signature of the $4\pl$ mass peak. A signal event is required to have a least two Z boson candidates reconstructed from the pairs of isolated
electrons or muons of opposite charges, and at least two jets which were identified as having originated from b quarks using multivariate (MVA) 
discriminants. The signal region is defined by requiring events to pass $115 <m_{4\pl}<135 $ GeV.
The dominant background is the SM single H production, followed by genuine nonresonant $\PZ\PZ^*$
events. To separate HH signal from the background a boosted decision tree (BDT) is trained using kinematic properties of signal events, the output of which is used in a multi-dimensional binned maximum likelihood fit to data to extract the results. Upper limits at 95\% CL are set on the signal strength of HH production with respect to the SM expectation at 30 (37), and $\kapl$ is constrained to be within the observed (expected)
range $-9~(-10.5) <\kapl < 14~(15.5)$ at 95\% CL. While the constraints from this channel are weaker than the results obtained from the combination of the other channels with partial Run-II dataset presented in Section~\ref{sec:overvew}, this clean final state will increasingly become more important with more data.  

\subsection{HH production via VBF in the $\bbbb$ final state}
The first result of the search for HH production via VBF in the $\bbbb$ final state is presented by the ATLAS Collaboration~\cite{Aad:2020kub}. The VBF production mode has unique sensitivity to the VVHH coupling modifier, $\k2v$, which controls the coupling strength with respect to its SM value. In addition, this analysis targets the resonant production mode, and searches for spin-0 resonances with masses in the range 260-1000 GeV considering broad-width (10-20\% of resonant mass) and narrow-width (4 MeV) hypotheses. The 4 b jets are tagged using MVA techniques, and the b jet energy resolution is improved by about 10\% with a dedicated b jet energy regression trained with a BDT. The main challenge in this search is an accurate estimation of the background dominated by multĳet production. The multijet background is modelled using data events with lower b jet multiplicity and reweighting them to model events with higher b jet multiplicity. The invariant mass of the HH system is reconstructed from the 4 b jets and is used as the final discriminant to extract the signal. The observed (expected) upper limits at 95\% CL are set on the cross section for nonresonant HH production via VBF at 840 (550) times the SM prediction, and also for resonant HH production via VBF as a function of resonance mass as shown on Fig.~\ref{fig:4b_res}. The 95\% CL observed (expected) allowed interval for the coupling modifier $\k2v$ is  $-0.43 < \k2v < 2.56$ ($-0.55 < \kapl < 2.72$). 

\begin{figure}[h!t]
\centering
\includegraphics[width=0.43\textwidth]{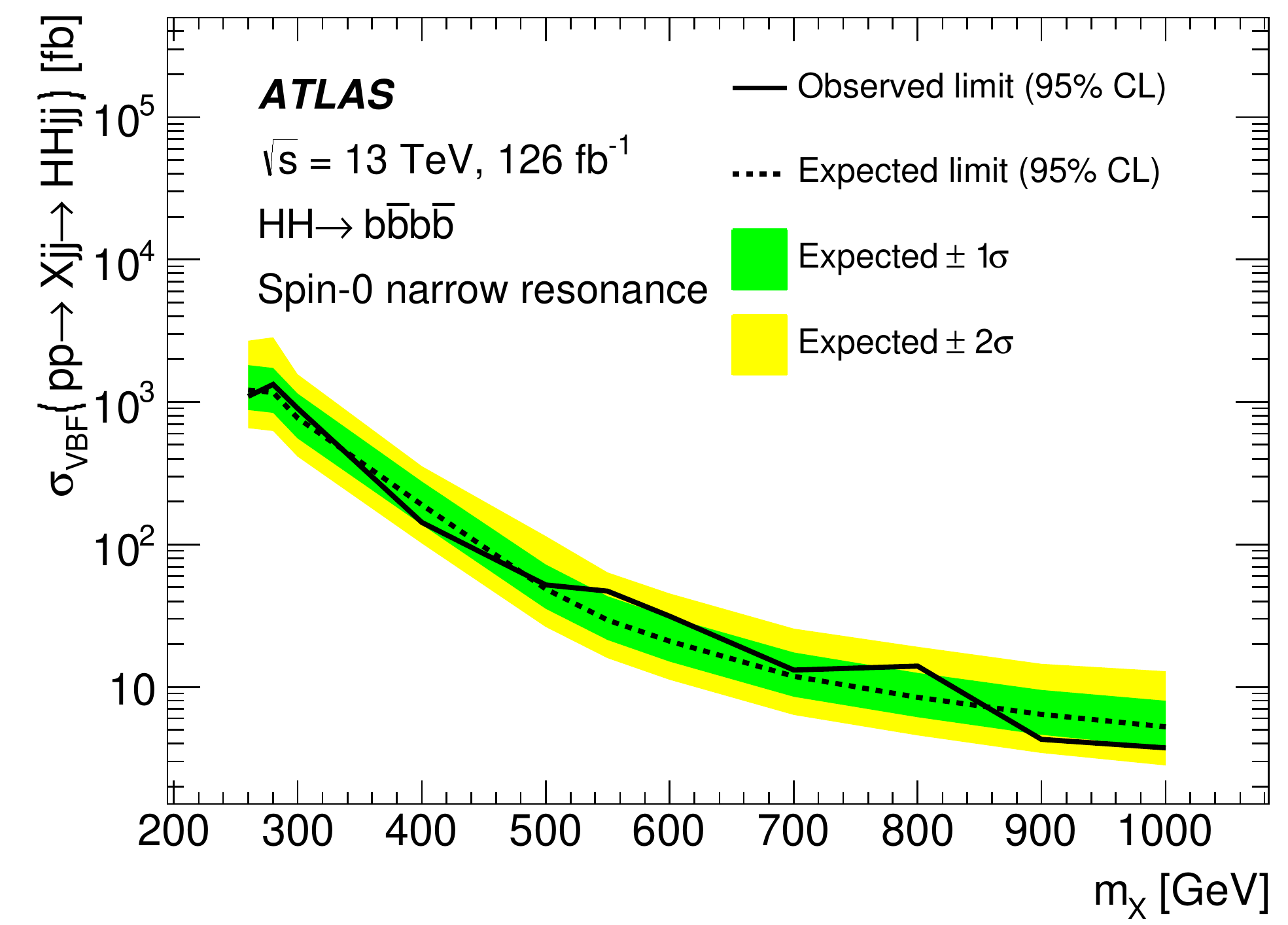}
\includegraphics[width=0.43\textwidth]{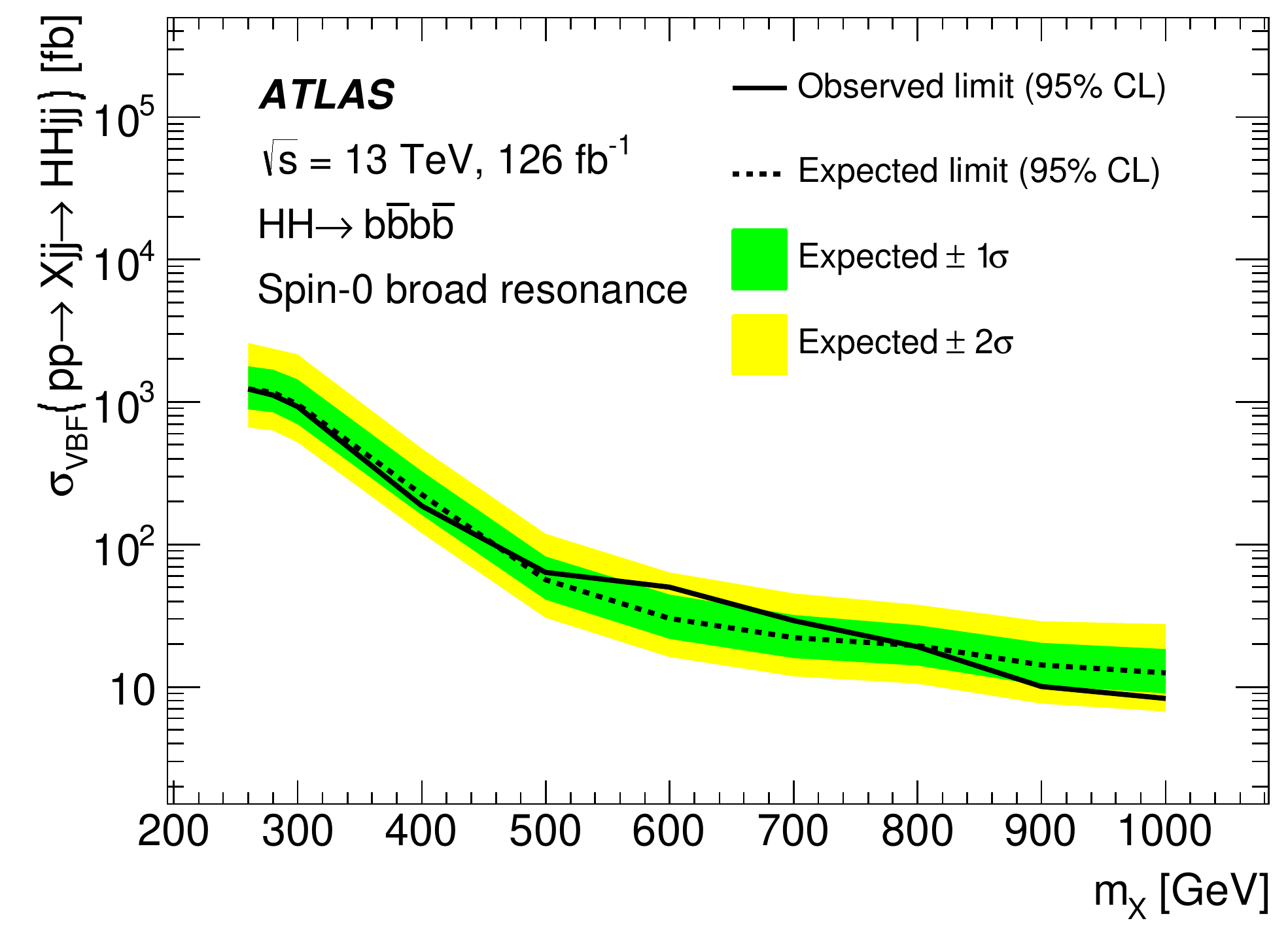}
\caption[Observed and expected upper limits at 95\% CL on the production cross section for resonant HH production via VBF as a function of resonant mass. The (a) narrow- and (b) broad-width resonance hypotheses are shown.]{\small Observed and expected upper limits at 95\% CL on the production cross section for resonant HH production via VBF as a function of resonant mass. The (a) narrow- and (b) broad-width resonance hypotheses are shown~\cite{Aad:2020kub}.}
\label{fig:4b_res}
\end{figure}

\subsection{Nonresonant HH production via ggF and VBF in the $\bbgg$ final state}
Search for nonresonant HH production via ggF and VBF in the final states with two photons and two bottom quarks is presented by the CMS Collaboration~\cite{Sirunyan:2020xok}. This final state has a tiny branching fraction of only 0.26\%, but offers a clean signature from Higgs boson decay to a pair of photons and excellent photon energy resolution. The energy resolution of b jets is improved by about 13\% by the deep neural network (DNN)-based b jet energy regression~\cite{Sirunyan:2019wwa}, which improves the dijet invariant mass of the SM HH signal by about 20\%. The main background is nonresonant $\ggbb$ production, followed by the single Higgs boson production in association with a top quark-antiquark pair ($\ttH$). The analysis is optimized to be sensitive to the SM HH production, anomalous values of $\kapl$, $\k2v$, and other beyond SM signals described by effective field theory. Both ggF and VBF production modes are analyzed following similar strategies: the background is suppressed using MVA techniques, and the VBF and ggF signal regions are defined based on the MVA purity and reconstructed invariant mass of the HH system. Finally, the signal is extracted from a fit to the invariant masses of the Higgs boson candidates in  the $\bbbar$ and $\gamma\gamma$ final states. The observed (expected) upper limit at 95\% CL on the HH production cross section  corresponds to 7.7 (5.2) times the SM prediction for the ggF mode and 225 (208) for the VBF mode . The observed (expected) constraints at 95\% CL on $\kapl$ and $\k2v$ are $-3.3 < \kapl < 8.5$ ($-2.5 < \kapl < 8.2)$ and $-1.3 < \k2v < 3.5$ ($-0.9 < \k2v < 3.1$) as shown in Fig.~\ref{fig:2b2g}. This search has 4 times better sensitivity than the previous analysis~\cite{Sirunyan:2018iwt} benefiting equally from larger data sample, and the innovative analysis techniques. In addition, the search was combined with an analysis that targets $\ttH$ where Higgs boson decays to a diphoton pair~\cite{Sirunyan:2020sum}, which allowed $\lbdHHH$ and top Yukawa coupling to be measured simultaneously, and provide constraints applicable to a wide range of theoretical models, where both couplings have anomalous values.

\begin{figure}[h!t]
\centering
\includegraphics[width=0.43\textwidth]{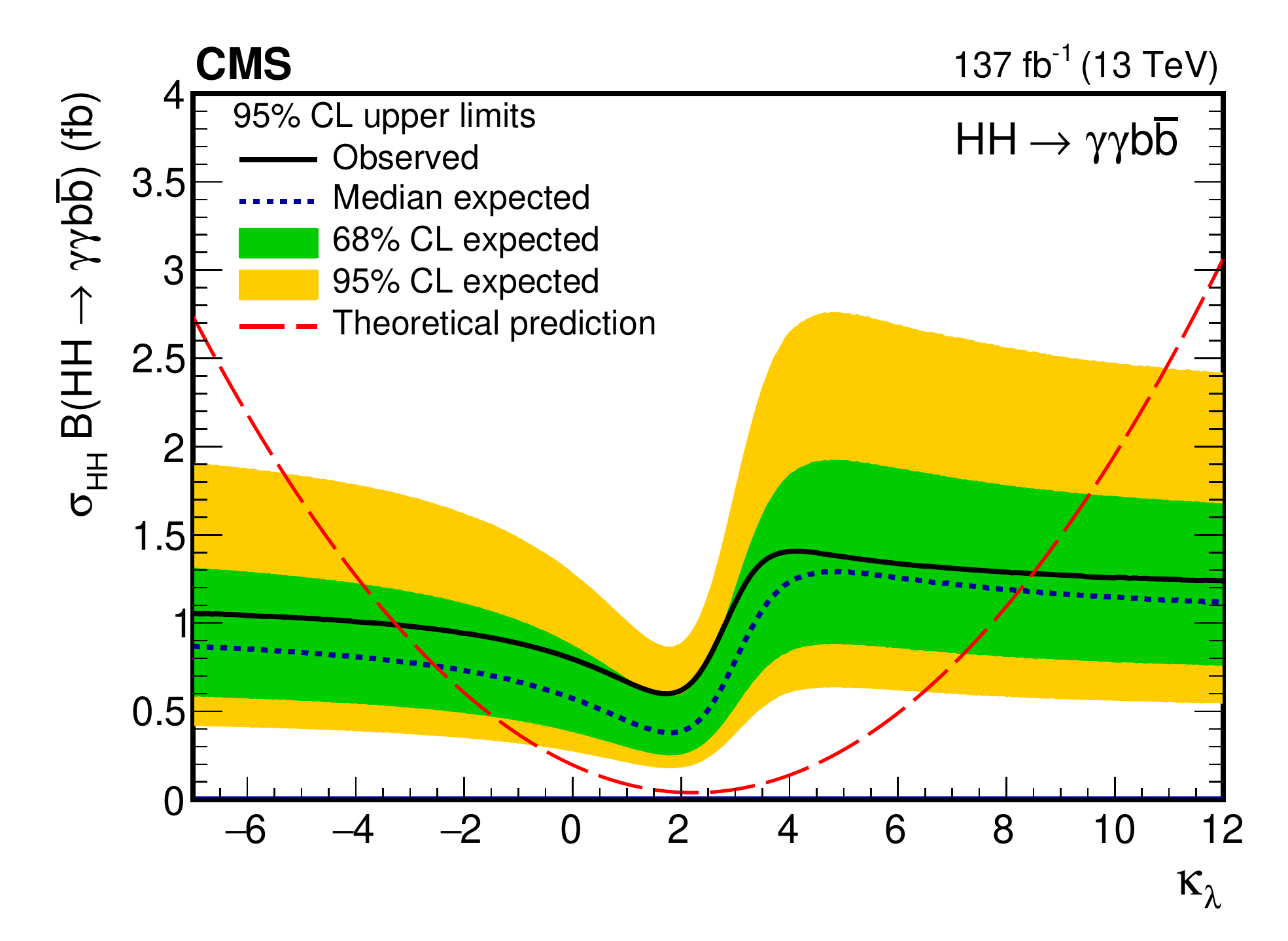}
\includegraphics[width=0.43\textwidth]{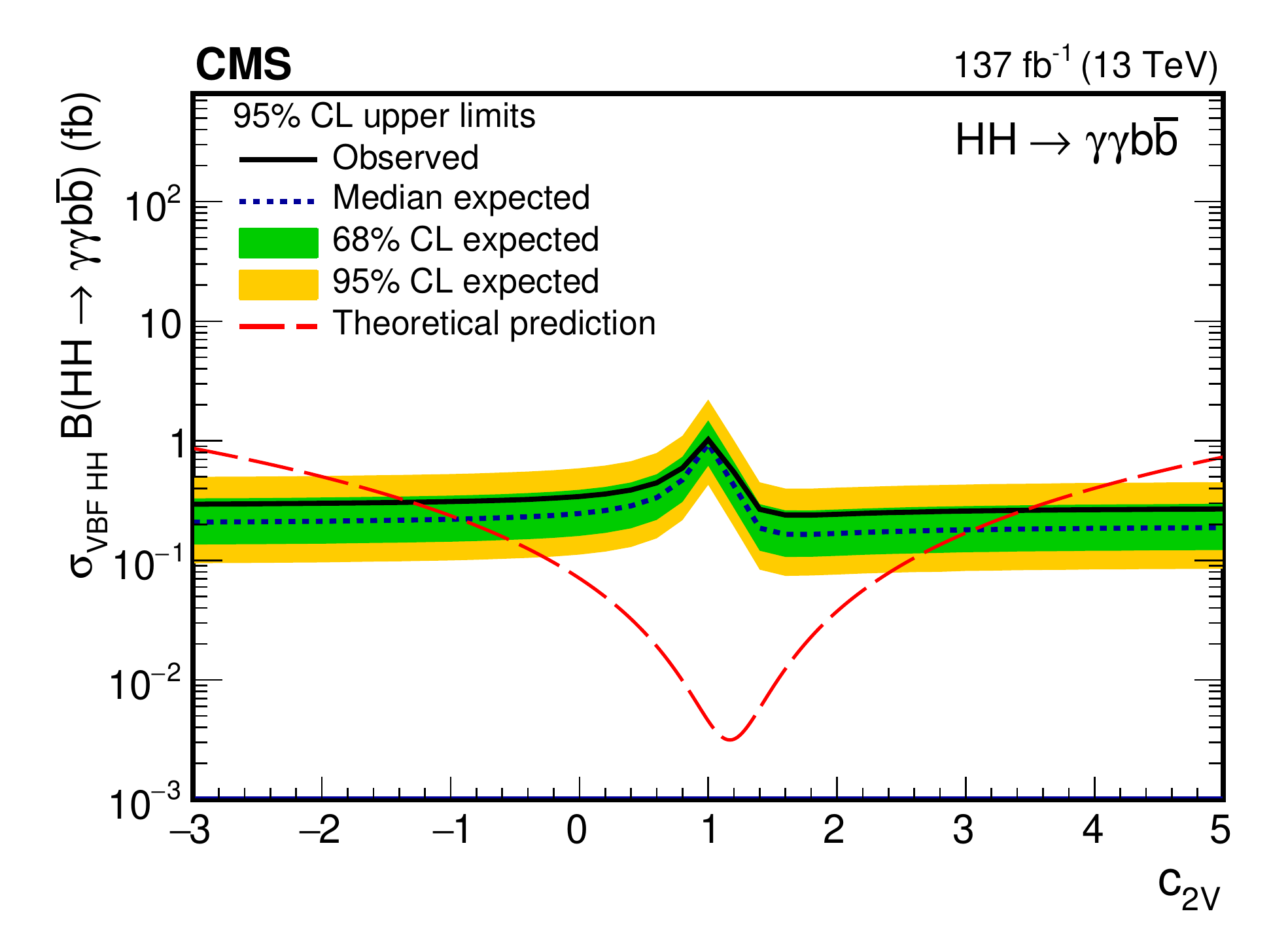}
\caption[Expected and observed 95\% CL upper limits on the product of the HH production cross section and branching fraction into the $\bbgg$ final state obtained for different values of $\kapl$ (left) and $\k2v$, denoted here as $\mathrm{c}_{2\PV}$, (right).]{\small  Expected and observed 95\% CL upper limits on the product of the HH production cross section and branching fraction into the $\bbgg$ final state obtained for different values of $\kapl$ (left) and $\k2v$, denoted here as $\mathrm{c}_{2\PV}$, (right)~\cite{Sirunyan:2020xok}.}
\label{fig:2b2g}
\end{figure}

\subsection{Resonant HH production in the boosted $\bbtt$ final state }
Search for a heavy, narrow, scalar resonance with mass in a range 1-3 TeV, produced via ggF and decaying to HH is presented by the ATLAS Collaboration~\cite{Aad:2020ldt}. The final state of interest is a boosted $\bbbar$ pair and a hadronically decaying boosted $\tau^{+}\tau^{-}$. A new technique, di-$\tau$ tagger, is developed to reconstruct boosted $\tau$ pair as a large radius jet with two sub-jets of smaller radii. For the di-$\tau$ tagger a BDT is employed to reject background of quark- and gluon-initiated jets, and the tagger efficiency is
measured in data. In the search for HH production the main background is from multi-jet production with misidentified di-$\tau$ objects, and $\PZ(\tau^{+}\tau^{-})$+jets production. To extract the HH signal a single-bin counting experiment
is performed for every considered resonance mass hypothesis. Upper limits at 95\% CL are set on HH production cross section via narrow width scalar resonance as shown in Fig.~\ref{fig:res}.

\begin{figure}[h!t]
\centering
\includegraphics[width=0.5\textwidth]{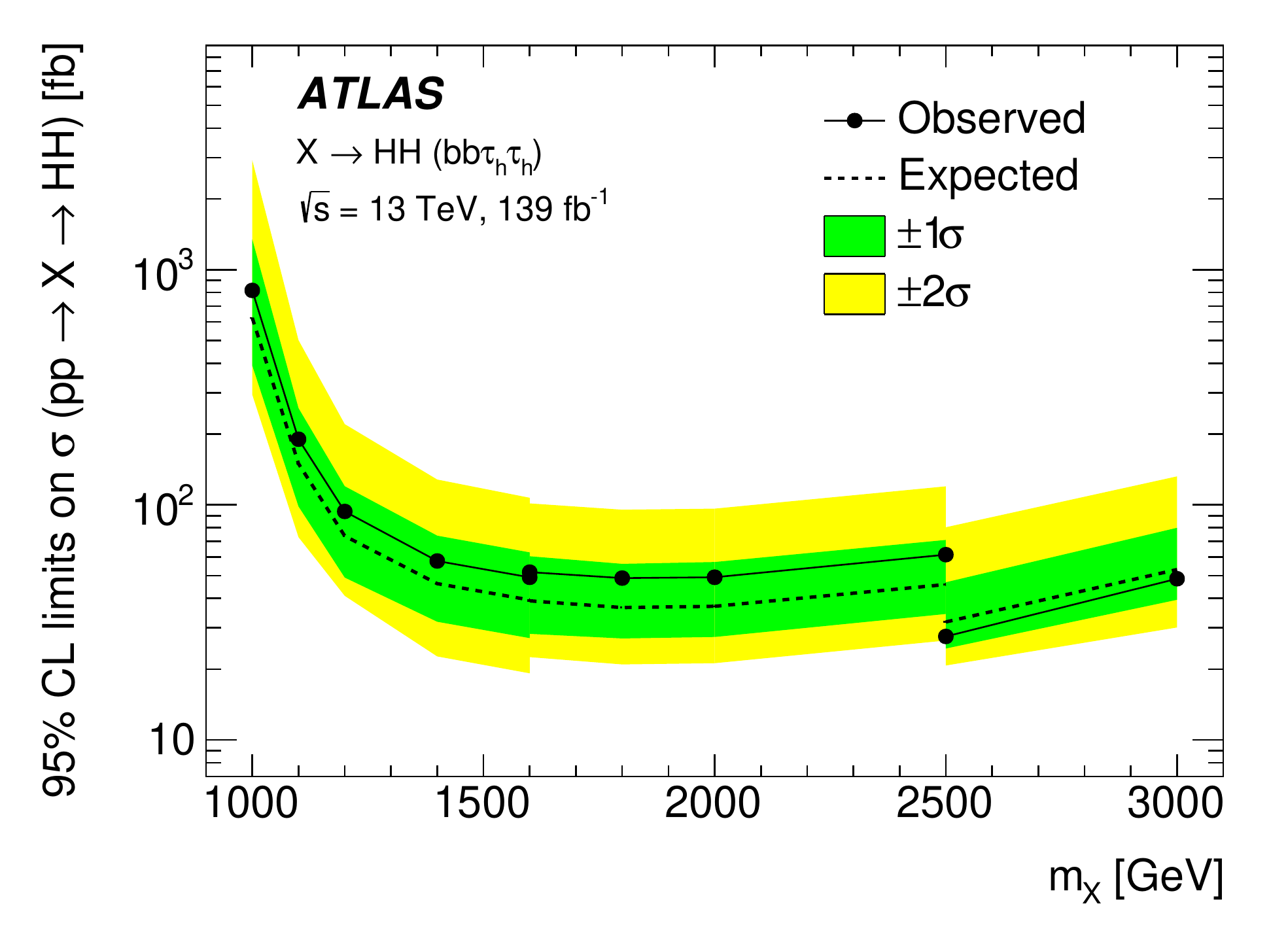}
\caption[Expected and observed 95\% CL upper limits on the production cross section of a heavy, narrow, scalar
resonance decaying into HH in the boosted $\bbtt$ final state.]{\small Expected and observed 95\% CL upper limits on the production cross section of a heavy, narrow, scalar
resonance decaying into HH in the boosted $\bbtt$ final state~\cite{Aad:2020ldt}. }
\label{fig:res}
\end{figure}

\subsection{Heavy Higgs boson decay into two lighter Higgs bosons in the $\bbtt$ final state}
A new search for the decay of a heavy resonance $\mathrm{X}$ into the Higgs boson and another resonance $\mathrm{Y}$ with a mass $\mathrm{m_Y} < \mathrm{m_X} - \mh$ is presented by the CMS Collaboration~\cite{CMS:2021hws}. It is motivated by the next-to-minimal supersymmetric SM model (NMSSM)~\cite{Ellwanger:2009dp}, and is the first search for this signature at the LHC. The branching fractions of the $\mathrm{Y}$ resonance
into SM particles are expected to be similar to the Higgs boson ones, and a promising signature of the Higgs boson decay into a pair of tau leptons and the decay of the $\mathrm{Y}$
into a pair of b quarks is considered. The dijet mas resolution is improved by the DNN-based b jet energy regression~\cite{Sirunyan:2019wwa}, the di-$\tau$ mass resolution is improved with a likelihood based method~\cite{Bianchini:2014vza}, and a kinematic fit to the $\bbtt$ system is used for each considered $\mathrm{m_Y}$ and $ \mathrm{m_X}$ mass hypothesis. A DNN multi-classifier is trained to separate the HH signal from the different types of background classified as events containing genuine $\tau$ pairs, top quark pairs, events with quark or gluon induced jets
misidentified as $\tau$, and other processes not included in the previous classes. The DNN output functions for background and signal classes are used in the maximum likelihood fit for the signal extraction. The search is performed in mass ranges of $\mathrm{m_X}$ $\in$ [240  GeV, 3 TeV] and $\mathrm{m_Y}$ $\in$ [60 GeV, 2.8 TeV]. No signal is observed in any of the investigated mass combinations and model-independent upper limits at 95\% CL are set on the production cross section $\mathrm{X} \rightarrow \mathrm{Y} \PH$ times branching fractions as shown in Fig.~\ref{fig:xyh}.

\begin{figure}[h!t]
\centering
\includegraphics[width=0.7\textwidth]{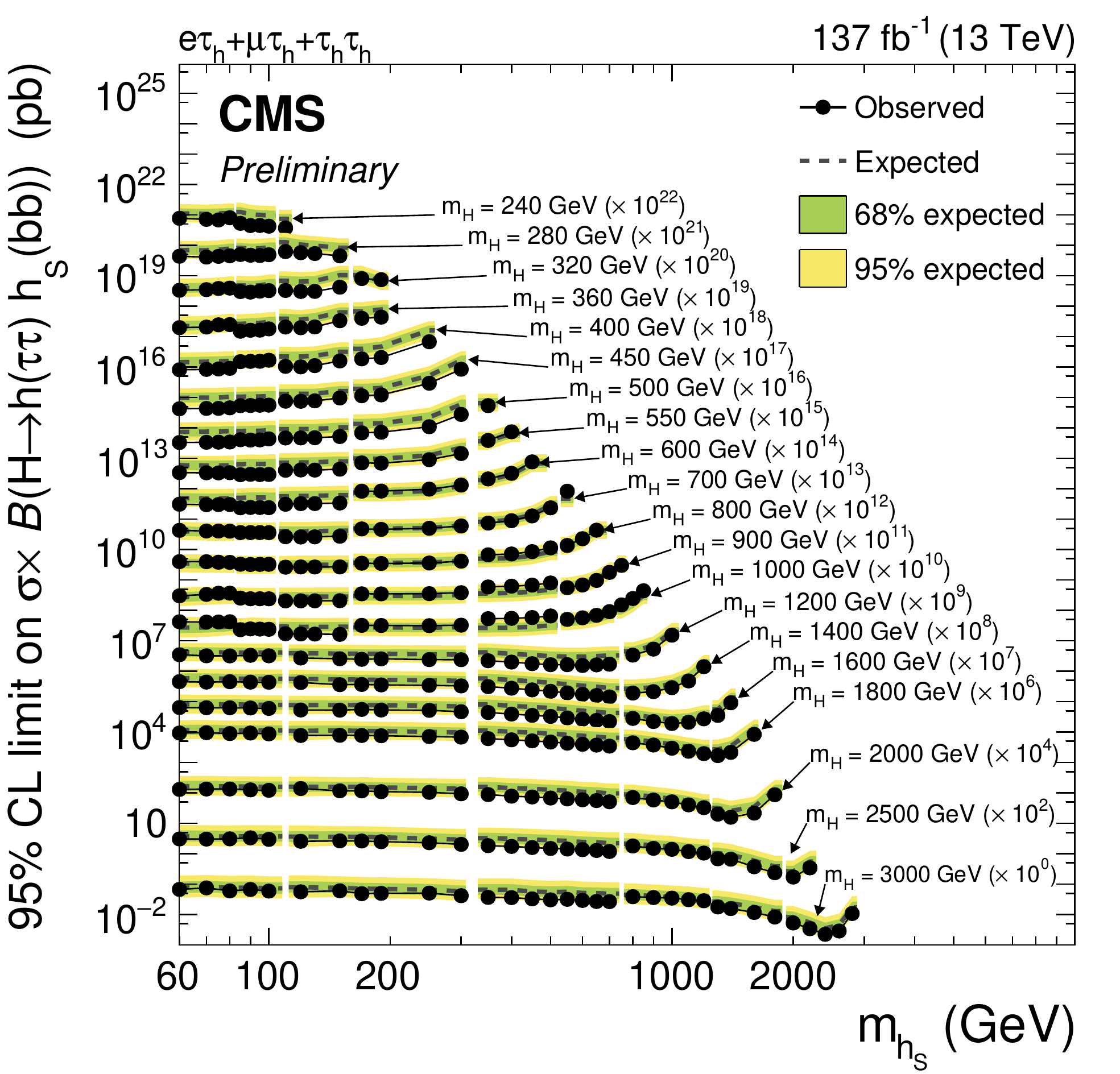}
\caption[Expected and observed 95\% CL upper limits on production cross section $\mathrm{X}\rightarrow \mathrm{Y} \PH$ times branching fractions for all
tested $\mathrm{m_Y}$ values. In this figure $\mathrm{X}\rightarrow \mathrm{Y} \PH$ is denoted as $\mathrm{H}\rightarrow \mathrm{h(\tau^{+}\tau^{-})} \mathrm{h_{S}}(\bbbar)$, where h is the observed Higgs boson, H is a heavy Higgs boson and $\mathrm{h_S}$ is another neutral Higgs boson, following the notations of NMSSM. The limits for each corresponding mass value have been scaled by orders of ten as indicated in the annotations.]{\small Expected and observed 95\% CL upper limits on production cross section $\mathrm{X}\rightarrow \mathrm{Y} \PH$ times branching fractions for all
tested $\mathrm{m_Y}$ values~\cite{CMS:2021hws}. In this figure $\mathrm{X}\rightarrow \mathrm{Y} \PH$ is denoted as $\mathrm{H}\rightarrow \mathrm{h(\tau^{+}\tau^{-})} \mathrm{h_{S}}(\bbbar)$, where h is the observed Higgs boson, H is a heavy Higgs boson and $\mathrm{h_S}$ is another neutral Higgs boson, following the notations of NMSSM. The limits for each corresponding mass value have been scaled by orders of ten as indicated in the annotations. }
\label{fig:xyh}
\end{figure}

\section{Future projections}
The first results from the di-Higgs searches using the full Run-II dataset presented in Section~\ref{sec:run2} showed significant analysis improvements. During the next Run-III of the LHC, which will begin in 2022, about 150 $\ifb$ of data are expected to be collected. With the improvements seen for the first results with the full Run-II dataset, when combining data from Run-II and Run-III we can expect to reach much better precision on HH production than was originally expected at the LHC. The next phase of LHC, the High Luminosity LHC (HL-LHC) is scheduled to start in 2027 and collect between 3000-4000 $\ifb$ of data with $\sqrt{s}=14$ TeV over more than a decade of operation. The HL-LHC physics prospects summary~\cite{Dainese:2019rgk} shows that the expected sensitivity of a combined ATLAS and CMS HH measurement is on the threshold of a discovery with the expected precision on $\kapl$ of 50\%. 

\section{Summary}
Searches for di-Higgs production at the LHC using the full Run-II dataset were presented. The new results benefited not only from the larger collected datasets, but also from a wealth of innovative analysis techniques, and showed exploration of rare channels and new signatures as well a probe of the VBF production mode and the anomalous VVHH coupling. Numerous beyond SM hypotheses and coupling modifications were explored in the context of resonant and nonresonant Higgs boson pair production. While all the results are so far consistent with the SM predictions, the exploration of the HH production at the LHC has just started. Many new HH results with the full Run-II data will follow in the near future, and we have very exciting prospects for the future data taking runs of the LHC.

\end{document}